\documentclass[twocolumn,amsmath,amssymb,10pt,aps]{revtex4}
  
\usepackage[utf8]{inputenc}
\usepackage[T1]{fontenc}

\usepackage{amsmath}
\usepackage{amsfonts}
\usepackage{graphicx}
\usepackage{yfonts}
\usepackage{color}
\usepackage[normalem]{ulem}
\usepackage{amsthm}
\usepackage{bm}
\usepackage{bbm}
\usepackage{mathtools}
\usepackage{array}
\usepackage{placeins}
\usepackage{enumitem}

\newcommand{\der}{\,\mathrm{d}}
\newcommand\bigforall{\mbox{\Large $\mathsurround=1pt\forall$}}

\def\<{\langle}
\def\>{\rangle}

\newcommand{\Tr}{\mathrm{Tr}}
\def\oper{{\mathchoice{\rm 1\mskip-4mu l}{\rm 1\mskip-4mu l}
{\rm 1\mskip-4.5mu l}{\rm 1\mskip-5mu l}}}
\DeclareMathAlphabet\mathbfcal{OMS}{cmsy}{b}{n}

\mathchardef\mhyphen="2D % Define a "math hyphen"

\newtheorem{Example}{Example}

\begin{document}

\title{Geometry of symmetric and non-invertible Pauli channels}

\author{Katarzyna Siudzi{\'n}ska}
\affiliation{Institute of Physics, Faculty of Physics, Astronomy and Informatics \\  Nicolaus Copernicus University, ul. Grudzi\k{a}dzka 5/7, 87--100 Toru{\'n}, Poland}

\begin{abstract}
We analyze the geometry of positive and completely positive, trace preserving Pauli maps that are fully determined by up to two distinct parameters. This includes five classes of symmetric and non-invertible Pauli channels. Using the Hilbert-Schmidt metric in the space of the Choi-Jamio{\l}kowski states, we compute the relative volumes of entanglement breaking, time-local generated, and divisible channels. Finally, we find the shapes of the complete positivity regions in relation to the tetrahedron of all Pauli channels.
\end{abstract}

\flushbottom

\maketitle

\thispagestyle{empty}

\section{Introduction}

Recently, much attention has been given to the geometrical
aspects of quantum channels. The questions that have been posed consider the shape and size of the space of channels, the properties of the underlying geometrical structures, or the total amount of quantum maps. Due to the calculations becoming more complex with the space dimension, the most is known about qubit maps.

The volume occupied by the channels that can be simulated by a single-qubit mixed-state environment has been computed for the generalized depolarizing channels \cite{Arvind} and the generalized amplitude damping channels \cite{Jung}. For the Pauli channels, a measure was proposed, according to which the volume of the positive but not completely positive, trace-preserving Pauli maps is twice the volume of the Pauli channels \cite{Jagadish}. Lovas and Andai \cite{Lovas} calculated the relative volume of unital qubit channels using the Lebesque measure.

For qudit case, Szarek et. al. \cite{Szarek2} provided the bounds for the volumes of positive, trace-preserving maps and the subsets of completely positive, decomposable, and superpositive qudit maps. Also, the relative volumes of the entanglement breaking and time-local generated generalized Pauli channels have recently been found \cite{GGPC_volume}. The geometry of the channels accessible via the Lindblad semigroup has been considered for both the Pauli channels \cite{Filippov,Karol_Pauli} and Weyl channels \cite{Karol_Weyl}. Additionally, for the Pauli channels, the volume of non-Markovian dynamical maps was found for the convex combinations of Markovian semigroups \cite{Jagadish2} and their simple generalization \cite{Jagadish3}. For continuous variable systems, the geometry of the Gaussian channels has been analyzed for the Bures-Fisher \cite{Monras2010} and Hilbert-Schmidt \cite{Gaussians} metrics.

This paper is a continuation of ref. \cite{Pauli_volume}, where we considered the geometry of the Pauli channels characterized by three distinct, non-zero parameters. Using the Hilbert-Schmidt metric on the space of the associated Choi-Jamio{\l}kowski states, the general forms of the line and volume elements were derived. By integrating the volume element over suitable regions, we computed the volumes of channels that are entanglement breaking, divisible, or obtainable through time-local generators.

This paper aims to answer the question to what extent the formalism of quantum channels and dynamical maps generated from master equations are compatible. For time-local master equations, this is quantified by the amount of the channels obtainable through time-local generators with respect to all channels. The volume expressions help us to learn more about the specific subsets of quantum channels. The volume ratios are often interpreted as probabilities that a randomly chosen channel possesses the requested properties.

In the forthcoming sections, we apply our earlier results to the Pauli channels characterized by no more than two distinct parameters. This class was completely omitted in ref. \cite{Pauli_volume}. It includes the non-invertible Pauli channels, where at least one of the three parameters vanishes, as well as the Pauli channels with additional symmetries, whose eigenvalues are degenerated. For each case, we present the integration regions and derive the volumes of positive and completely positive, trace-preserving Pauli maps. Next, we analyze and compare the relative volumes between different classes of the Pauli maps. We illustrate our results by plotting the crossections of their complete positivity regions with the tetrahedron of all Pauli channels. In addition, we demonstrate the volume ratios for Pauli maps and divisible Pauli channels in two collective charts.

\section{Pauli channels and dynamical maps}

Consider the trace-preserving Pauli map defined by
\begin{equation}\label{Pauli}
\Lambda[X]=\sum_{\alpha=0}^3p_\alpha\sigma_\alpha X\sigma_\alpha
\end{equation}
with real $p_\alpha$ such that $\sum_{\alpha=0}^3p_\alpha=1$ and the Pauli matrices $\sigma_\alpha$. Alternatively, this map can be given via the eigenvalue equations
\begin{equation}
\Lambda[\sigma_\alpha]=\lambda_\alpha\sigma_\alpha,
\end{equation}
where the eigenvalues $\lambda_\alpha$ are related to $p_\alpha$ as follows,
\begin{equation}
\lambda_0=1,\qquad\lambda_k=2(p_0+p_k)-1.
\end{equation}
Now, $\Lambda$ preserves the positivity of input states provided that \cite{Zyczkowski,Jagadish}
\begin{equation}\label{P}
|\lambda_k|\leq 1,
\end{equation}
which is the necessary and sufficient condition for its positivity.
Moreover, $\Lambda$ is a quantum channel (completely positive, trace-preserving map) if and only if \cite{Fujiwara,Ruskai}
\begin{equation}\label{CP}
|1\pm\lambda_3|\geq|\lambda_1\pm\lambda_2|.
\end{equation}
A special class of the Pauli channels consists in the entanglement breaking channels. These are the channels that, when extended to $\oper_N\otimes\Lambda$ with any $N$-dimensional identity map, always produce separable output states. 
The necessary and sufficient condition for the Pauli channel to break quantum entanglement is given by \cite{qubitEBC}
\begin{equation}\label{EB}
\sum_{k=1}^3|\lambda_k|\leq 1.
\end{equation}

Quantum channels are usually used to describe discrete changes of a physical system. Examples include quantum processing, noise effects on quantum register, and quantum measurement \cite{TQI}. However, there is a way to model continuous time-evolution using the families $\{\Lambda(t)|t\geq 0,\Lambda(0)=\oper\}$ of time-parameterized quantum channels $\Lambda(t)$, known as {\it dynamical maps}.
The simplest example of the Pauli dynamical map is the Markovian semigroup $\Lambda(t)$, which solves the master equation
\begin{equation}\label{MS}
\dot{\Lambda}(t)=\mathcal{L}\Lambda(t),\qquad\Lambda(0)=\oper.
\end{equation}
The time-independent generator $\mathcal{L}$ has the Gorini-Kossakowski-Sudarshan-Lindblad (GKSL) form \cite{GKS,L}
\begin{equation}\label{L}
\mathcal{L}[\rho]=\frac 12 \sum_{k=1}^3\gamma_k(\sigma_k\rho \sigma_k-\rho),
\end{equation}
where $\gamma_k\geq 0$. Non-Markovian effects of quantum evolution are included via the time-local generators $\mathcal{L}(t)$. They have the GKSL form, but their decoherence rates $\gamma_k(t)$ no longer have to be positive. The eigenvalues of the corresponding dynamical map $\Lambda(t)$ read \cite{mub_final}
\begin{equation}
\lambda_k(t)=\exp\left[-\sum_{j\neq k}\int_0^t\gamma_j(\tau)\der\tau\right].
\end{equation}
It is important to note that a Pauli channel $\Lambda$ is obtainable from the Pauli dynamical map $\Lambda(t)$ by fixing the time $t=t_\ast\geq 0$. However, this is possible if and only if
\begin{equation}\label{TLG}
\bigforall_{k=1,2,3}\quad\lambda_k\geq 0.
\end{equation}
We call these channels {\it obtainable from time-local generators}.

\section{Volume of Pauli channels}

Consider the geometry of Pauli maps $\Lambda$ in the subspace of state space defined by the Choi-Jamio{\l}kowski isomorphism \cite{Choi,Jamiolkowski}
\begin{equation}
\rho_\Lambda=\frac 12 \sum_{i,j=0}^1|i\>\<j|\otimes\Lambda[|i\>\<j|].
\end{equation}
We equip this space with the Hilbert-Schmidt metric $g=\frac 14\mathrm{diag}(1,1,1)$ induced by the line element $\der s^2:=\Tr(\der\rho_\Lambda)^2$. For general Pauli maps from eq. (\ref{Pauli}), one has \cite{Pauli_volume}
\begin{equation}\label{ds2}
\der s^2=\frac 14 (\der\lambda_1^2+\der\lambda_2^2+\der\lambda_3^2),
\end{equation}
as well as the volume element
\begin{equation}\label{dV}
\der V:=\sqrt{\det g}\der\lambda_1\der\lambda_2\der\lambda_3=\frac 18 \der\lambda_1\der\lambda_2\der\lambda_3.
\end{equation}
The volumes of Pauli maps and Pauli channels
\begin{equation}
V(\mathcal{C}_{\mathrm{M}})=\int_{\mathcal{C}_{\mathrm{M}}}\der V
\end{equation}
are obtained via intergation over the regions $\mathcal{C}_{\mathrm{M}}$; in particular,
\begin{itemize}
\item the positivity region from eq. (\ref{P}),
\begin{equation}
\mathcal{C}_{\mathrm{PT}}:=\Big\{\lambda_k:\,|\lambda_k|\leq 1,\,k=1,2,3\Big\},
\end{equation}
\item the complete positivity region determined by eq. (\ref{CP}),
\begin{equation}
\mathcal{C}_{\mathrm{CPT}}=\Big\{\lambda_1,\lambda_2,\lambda_3:\,|1\pm\lambda_3|\geq|\lambda_1\pm\lambda_2|\Big\},
\end{equation}
\item the region of entanglement breaking channels -- see eq. (\ref{EB}),
\begin{equation}
\mathcal{C}_{\mathrm{EBC}}=\left\{\lambda_1,\lambda_2,\lambda_3:\,\sum_{k=1}^3|\lambda_k|\leq 1\right\},
\end{equation}
\item the region given by eq. (\ref{TLG}) that consists in the maps obtainable from time-local generators,
\begin{equation}
\mathcal{C}_{\mathrm{TLG}}=\Big\{\lambda_k:\,\lambda_k\geq 0,\,k=1,2,3\Big\}.
\end{equation}
\end{itemize}

While the formula for the line element is universal, the volume element changes when one considers special families of the Pauli maps. For example, in the general case, the channels with even a single $\lambda_k=0$ have no input into the volume, and yet relative volumes for such maps can still be computed. In the following subsections, we restrict our attention to the subspaces of symmetric (with degenerated eigenvalues) or non-invertible (with vanishing eigenvalues) Pauli channels, and we analyze their geometrical properties.

\subsection{One-parameter channels}

\begin{figure}[htb!]
  \includegraphics[width=0.3\textwidth]{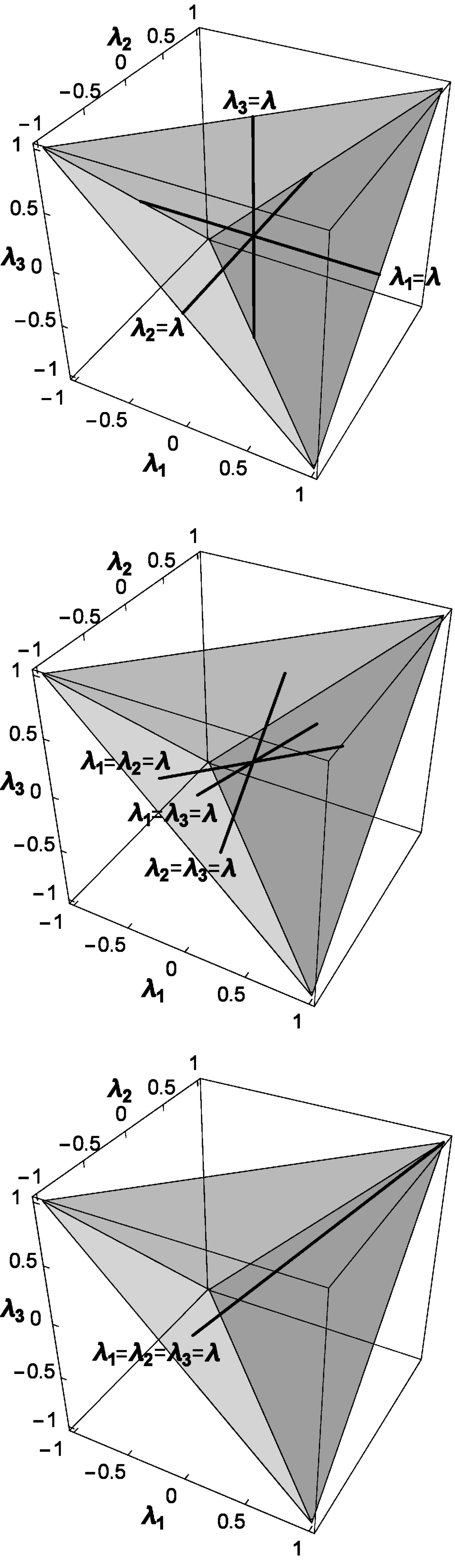}
\caption{
A graphical representation of the range of eigenvalues $\lambda_1,\ \lambda_2,\ \lambda_3$ that correspond to the one-parameter channels (bold line segments) with one (top), two (middle), and three (bottom) identical non-zero eigenvalues. The gray regions are the faces of the tetrahedron of all Pauli channels.}
\label{one}
\end{figure}

First, let us consider the Pauli channels that are fully-determined by one real parameter $\lambda\neq 0$. Such channels occupy the line segments that lie inside the tetrahendron of all Pauli channels (see Fig.~\ref{one}). The underlying manifolds are one-dimensional, hence the associated volumes are simply the lengths of the line segments.

\FloatBarrier

The easiest example of a one-parameter channel corresponds to the choice
\begin{equation}
\{\lambda_1,\lambda_2,\lambda_3\}=\{\lambda,0,0\},
\end{equation}
where there is only one non-zero eigenvalue $\lambda$. These channels are known in literature as {\it axial} or {\it linear} \cite{Imre}. Evidently, they are non-invertible, as two of the eigenvalues are always equal to zero. One finds
\begin{equation}\label{V1}
\der s^2=\frac 14 \der\lambda^2,\qquad \der V=\frac 12 \der\lambda,
\end{equation}
and also
\begin{align*}
\mathcal{C}_{\mathrm{PT}}=\mathcal{C}_{\mathrm{CPT}}=\mathcal{C}_{\mathrm{EBC}}
&=\Big\{\lambda:\,-1\leq\lambda\leq 1\Big\},
\\
\mathcal{C}_{\mathrm{PT}}\cap\mathcal{C}_{\mathrm{TLG}}&=\Big\{\lambda:\,0\leq\lambda\leq 1\Big\}.
\end{align*}
Hence, every positive, trace-preserving map is automatically completely positive and entanglement breaking. Moreover, $V(\mathcal{C}_{\mathrm{PT}})=1$ and
\begin{equation}
\frac{V(\mathcal{C}_{\mathrm{PT}}\cap\mathcal{C}_{\mathrm{TLG}})}{V(\mathcal{C}_{\mathrm{PT}})}
%=\displaystyle\frac{\frac 12 \int_{0}^1\der\lambda}{\frac 12 \int_{-1}^1\der\lambda}
=\frac 12,
\end{equation}
so exactly half of these maps can be obtained with time-local generators.

A more complicated case is when there are two identical non-zero eigenvalues $\lambda$,
\begin{equation}
\{\lambda_1,\lambda_2,\lambda_3\}=\{\lambda,\lambda,0\}.
\end{equation}
Such channels are both non-invertible and symmetric due to $\lambda$ being two-times degenerate. The line and volume elements
\begin{equation}\label{V2}
\der s^2=\frac 12 \der\lambda^2,\qquad \der V=\frac{\sqrt{2}}{2} \der\lambda
\end{equation}
differ from the ones in eq. (\ref{V1}) only by a normalization factor. However, there is a greater variety among the integration regions. Namely,
\begin{align*}
\mathcal{C}_{\mathrm{PT}}&=\Big\{\lambda:\,-1\leq\lambda\leq 1\Big\},
\\\mathcal{C}_{\mathrm{CPT}}=\mathcal{C}_{\mathrm{EBC}}
&=\Big\{\lambda:\,-1/2\leq\lambda\leq 1/2\Big\},
\\
\mathcal{C}_{\mathrm{PT}}\cap\mathcal{C}_{\mathrm{TLG}}
&=\Big\{\lambda:\,0\leq\lambda\leq 1\Big\},
\\
\mathcal{C}_{\mathrm{CPT}}\cap\mathcal{C}_{\mathrm{TLG}}
&=\Big\{\lambda:\,0\leq\lambda\leq 1/2\Big\}.
\end{align*}
Note that this time only half of the positive, trace-preserving maps are completely positive, where $V(\mathcal{C}_{\mathrm{PT}})=\sqrt{2}$,
and all the channels are entanglement breaking. Insterestingly, there are as many positive, trace-preserving maps obtainable with time-local generators as there are quantum channels,
\begin{equation}
V(\mathcal{C}_{\mathrm{PT}}\cap\mathcal{C}_{\mathrm{TLG}})=V(\mathcal{C}_{\mathrm{CPT}})=\frac{\sqrt{2}}{2}.
\end{equation}
Again, only half of the maps that belong to $\mathcal{C}_{\mathrm{PT}}\cap\mathcal{C}_{\mathrm{TLG}}$ are completely positive. Finally, observe that there are no Pauli dynamical maps that always have two identical non-zero eigenvalues, as $\lambda(0)=1\notin\mathcal{C}_{\mathrm{CPT}}$.

Lastly, one can also have three identical non-zero eigenvalues $\lambda$,
\begin{equation}
\{\lambda_1,\lambda_2,\lambda_3\}=\{\lambda,\lambda,\lambda\}.
\end{equation}
The channel whose all eigenvalues are equal is called the {\it depolarizing channel}, and $\lambda$ is the degree of depolarization. They are highly symmetric and the only invertible maps in this subsection. One has
\begin{equation}
\der s^2=\frac 34 \der\lambda^2,\qquad\der V=\frac{\sqrt{3}}{2}\der\lambda.
\end{equation}
It is easy to check that
\begin{align*}
\mathcal{C}_{\mathrm{PT}}&=\Big\{\lambda:\,-1\leq\lambda\leq 1\Big\},
\\\mathcal{C}_{\mathrm{CPT}}=\mathcal{C}_{\mathrm{EBC}}
&=\Big\{\lambda:\,-1/3\leq\lambda\leq 1\Big\},
\\
\mathcal{C}_{\mathrm{PT}}\cap\mathcal{C}_{\mathrm{TLG}}
&=\mathcal{C}_{\mathrm{CPT}}\cap\mathcal{C}_{\mathrm{TLG}}\\
&=\Big\{\lambda:\,0\leq\lambda\leq 1\Big\}.
\end{align*}
Now, the volumes of all positive and completely positive, trace-preserving Pauli maps are
\begin{equation}
V(\mathcal{C}_{\mathrm{PT}})=\sqrt{3},\qquad V(\mathcal{C}_{\mathrm{CPT}})=\frac{2\sqrt{3}}{3}.
\end{equation}
Only half of the Pauli channels break quantum entanglement, and just another half of them can be generated by time-local generators. The volume ratios for the maps with positive eigenvalues differ,
\begin{equation}
\frac{V(\mathcal{C}_{\mathrm{PT}}\cap\mathcal{C}_{\mathrm{TLG}})}
{V(\mathcal{C}_{\mathrm{PT}})}=\frac 12,
\qquad
\frac{V(\mathcal{C}_{\mathrm{CPT}}\cap\mathcal{C}_{\mathrm{TLG}})}
{V(\mathcal{C}_{\mathrm{CPT}})}=\frac 34,
\end{equation}
even though $V(\mathcal{C}_{\mathrm{PT}}\cap\mathcal{C}_{\mathrm{TLG}})=
V(\mathcal{C}_{\mathrm{CPT}}\cap\mathcal{C}_{\mathrm{TLG}})$.

To summarize, the volume of all one-parameter Pauli maps that are positive and trace-preserving is different for every distinct class. This leads to the volumes of Pauli channels and their subclasses being unequal, as well. However, there are some similarities shared by all one-parameter maps. Namely, half of the positive, trace-preserving maps are obtainable with time-local generators, 
and every channel is entanglement breaking. A more detailed comparison is presented in Fig. \ref{CPT_chart}.

\FloatBarrier

\subsection{Two-parameter channels}

\FloatBarrier

\begin{figure}[htb!]
  \includegraphics[width=0.3\textwidth]{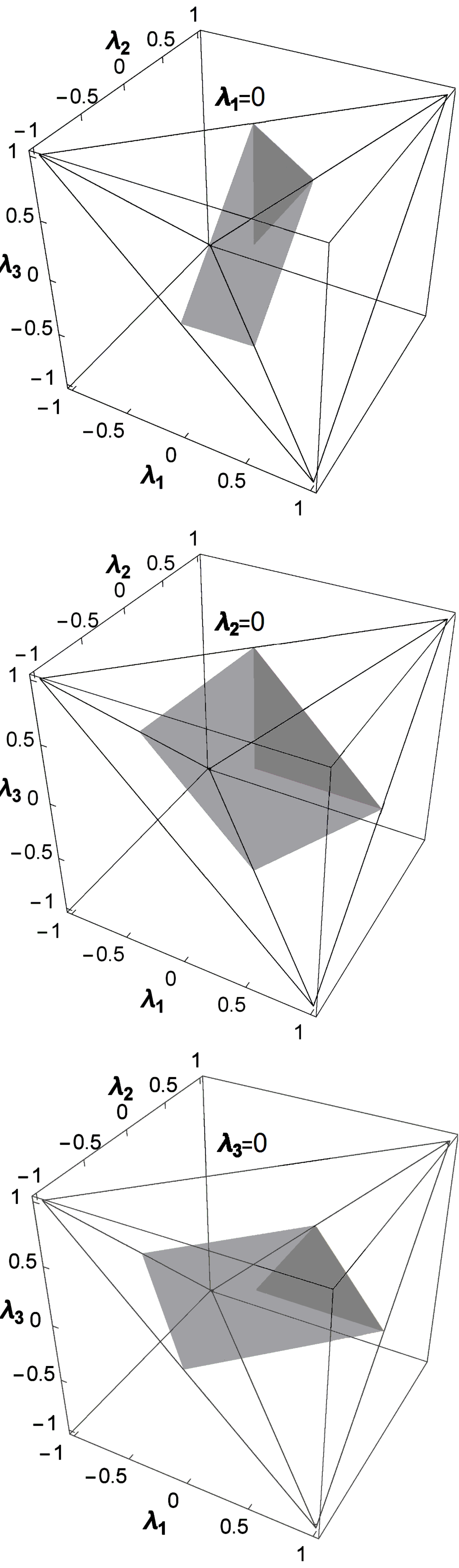}
\caption{
A graphical representation of the range of eigenvalues $\lambda_1,\ \lambda_2,\ \lambda_3$ that correspond to the two-parameter channels (gray flat surfaces) with two distinct non-zero eigenvalues. From top to bottom, these are the complete positivity regions for $\lambda_1=0$, $\lambda_2=0$, and $\lambda_3=0$, respectively. The dark gray parts correspond to the channels obtainable with time-local generators.}
\label{two}
\end{figure}

Now, we analyze the Pauli channels defined by two real non-zero parameters $\lambda\neq\eta$. Let us start with the maps that have two distinct non-zero eigenvalues,
\begin{equation}
\{\lambda_1,\lambda_2,\lambda_3\}=\{\lambda,\eta,0\}.
\end{equation}
The channels defined by the above choice of $\lambda_k$ are both non-symmetric and non-invertible. As shown in Fig.~\ref{two}, they live in crossections of flat surfaces with the tetrahendron of all Pauli channels. Note that the complete positivity regions are squares with edges lying on the faces of the tetrahedron. One finds
\begin{equation}
\der s^2=\frac 14 (\der\lambda^2+\der\eta^2),\qquad\der V=\frac 14 \der\lambda\der\eta,
\end{equation}
along with
\begin{align*}
\mathcal{C}_{\mathrm{PT}}&=\Big\{\lambda,\eta:\,-1\leq\lambda\leq 1,\,-1\leq\eta\leq 1\Big\},
\\\mathcal{C}_{\mathrm{CPT}}=\mathcal{C}_{\mathrm{EBC}}
&=\Big\{\lambda,\eta:\,-1\leq\lambda\leq 1,\\
&\qquad\qquad\,\,-1+|\lambda|\leq\eta\leq 1-|\lambda|\Big\},\\
\mathcal{C}_{\mathrm{PT}}\cap\mathcal{C}_{\mathrm{TLG}}
&=\Big\{\lambda,\eta:\,0\leq\lambda\leq 1,\,0\leq\eta\leq 1\Big\},
\\
\mathcal{C}_{\mathrm{CPT}}\cap\mathcal{C}_{\mathrm{TLG}}
&=\Big\{\lambda,\eta:\,0\leq\lambda\leq 1,\,0\leq\eta\leq 1-\lambda\Big\}.
\end{align*}
The volume of positive, trace-preserving maps is $V(\mathcal{C}_{\mathrm{PT}})=1$, which is exactly twice the volume of completely positive, trace-preserving maps. In this family of two-parameter maps, every Pauli channel is entanglement breaking. Finally, the relative volumes of positive and completely positive maps that are obtainable with time-local generators are both the same and equal to
\begin{equation}
\frac{V(\mathcal{C}_{\mathrm{PT}}\cap\mathcal{C}_{\mathrm{TLG}})}{V(\mathcal{C}_{\mathrm{PT}})}=
\frac{V(\mathcal{C}_{\mathrm{CPT}}\cap\mathcal{C}_{\mathrm{TLG}})}{V(\mathcal{C}_{\mathrm{CPT}})}
=\frac 14.
\end{equation}

\FloatBarrier

\begin{figure}[htb!]
  \includegraphics[width=0.3\textwidth]{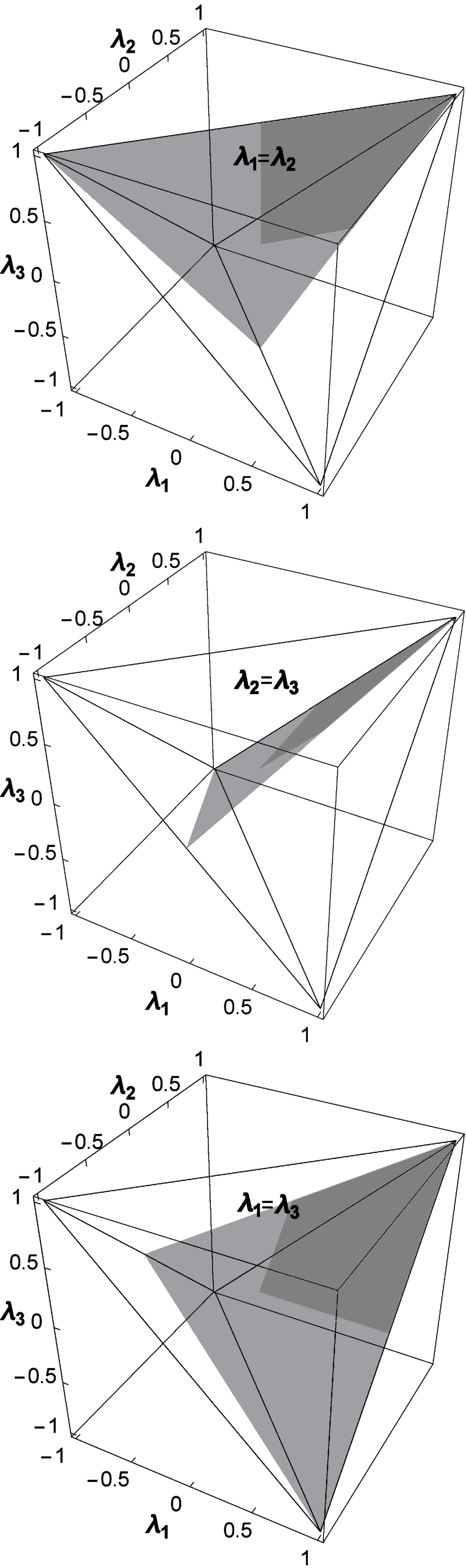}
\caption{
A graphical representation of the range of eigenvalues $\lambda_1,\ \lambda_2,\ \lambda_3$ that correspond to the two-parameter channels (gray flat surfaces) with two identical and one different non-zero eigenvalues. From top to bottom, these are the complete positivity regions for $\lambda_1=\lambda_2$, $\lambda_2=\lambda_3$, and $\lambda_3=\lambda_1$, respectively. The dark gray parts correspond to the channels obtainable with time-local generators.}
\label{three}
\end{figure}

The final example are the channels with two identical and one different non-zero eigenvalues,
\begin{equation}\label{Lam}
\{\lambda_1,\lambda_2,\lambda_3\}=\{\lambda,\lambda,\eta\}.
\end{equation}
It is easy to see that they are symmetric and invertible. In Fig.~\ref{three}, we see that the complete positivity regions are isosceles triangles with bases collinear with the edges of the tetrahedron. The line and volume elements read
\begin{equation}
\der s^2=\frac 14 (2\der\lambda^2+\der\eta^2),\qquad\der V=\frac{\sqrt{2}}{4} \der\lambda\der\eta.
\end{equation}
The integration regions are even more involved,
\begin{align*}
\mathcal{C}_{\mathrm{PT}}&=\Big\{\lambda,\eta:\,-1\leq\lambda\leq 1,\,-1\leq\eta\leq 1\Big\},\\
\mathcal{C}_{\mathrm{CPT}}&=\Big\{\lambda,\eta:\,-1\leq\eta\leq 1,\\
&\qquad\qquad\,\,-(1+\eta)/2\leq\lambda\leq(1+\eta)/2\Big\},\\
\mathcal{C}_{\mathrm{EBC}}&=\Big\{\lambda,\eta:\,-1\leq\eta\leq 1
,\\&\qquad\qquad\,\,-(1-|\eta|)/2\leq\lambda\leq(1-|\eta|)/2\Big\},\\
\mathcal{C}_{\mathrm{PT}}\cap\mathcal{C}_{\mathrm{TLG}}
&=\Big\{\lambda,\eta:\,0\leq\lambda\leq 1,\,0\leq\eta\leq 1\Big\},
\\
\mathcal{C}_{\mathrm{CPT}}\cap\mathcal{C}_{\mathrm{TLG}}
&=\Big\{\lambda,\eta:\,0\leq\eta\leq 1,\,0\leq\lambda\leq(1+\eta)/2\Big\},\\
\mathcal{C}_{\mathrm{EBC}}\cap\mathcal{C}_{\mathrm{TLG}}
&=\Big\{\lambda,\eta:\,0\leq\eta\leq 1,\,0\leq\lambda\leq(1-\eta)/2\Big\},\\
\end{align*}
Observe that this is the only family of one or two-parameter Pauli channels where not every channel breaks quantum entanglement. Indeed, there is an equal amount of the channels that are and are not entanglement breaking, as
\begin{equation}
\frac{V(\mathcal{C}_{\mathrm{EBC}})}{V(\mathcal{C}_{\mathrm{CPT}})}=\frac 12.
\end{equation}
Meanwhile, half of the positive, trace-preserving Pauli maps are also quantum channels, where one has $V(\mathcal{C}_{\mathrm{PT}})=\sqrt{2}$. As for the maps obtainable with time-local generators, the relative volumes read
\begin{equation}
\frac{\mathcal{C}_{\mathrm{PT}}\cap\mathcal{C}_{\mathrm{TLG}}}{\mathcal{C}_{\mathrm{PT}}}=
\frac{\mathcal{C}_{\mathrm{EBC}}\cap\mathcal{C}_{\mathrm{TLG}}}{\mathcal{C}_{\mathrm{EBC}}}
=\frac 14
\end{equation}
and
\begin{equation}
\frac{\mathcal{C}_{\mathrm{CPT}}\cap\mathcal{C}_{\mathrm{TLG}}}{\mathcal{C}_{\mathrm{CPT}}}
=\frac 38.
\end{equation}

The well-known channels that belong to the class given by eq. (\ref{Lam}) are the two-Pauli channels \cite{Fuchs} with
\begin{equation}
\{\lambda_1,\lambda_2,\lambda_3\}=\{1-\lambda,1-\lambda,1-2\lambda\},
\end{equation}
as well as the bit flip, phase flip (phase damping, dephasing), and bit-phase flip channels \cite{Imre} represented by
\begin{equation}
\{\lambda_1,\lambda_2,\lambda_3\}=\{1,1-2\lambda,1-2\lambda\}.
\end{equation}
Both of them are the one-parameter maps with
\begin{equation}
\mathcal{C}_{\mathrm{PT}}=\mathcal{C}_{\mathrm{CPT}}=
\Big\{\lambda:\,0\leq\lambda\leq 1\Big\}
\end{equation}
and
\begin{equation}
\mathcal{C}_{\mathrm{CPT}}\cap\mathcal{C}_{\mathrm{TLG}}=
\Big\{\lambda:\,0\leq\lambda\leq 1/2\Big\}.
\end{equation}
Therefore, all positive, trace-preserving dephasing and two-Pauli maps are completely positive, half of which are generatable with time-local generators. However, their entanglement breaking properties differ. For dephasing channels, one finds that $\mathcal{C}_{\mathrm{EBC}}=\{\lambda:\,1/2\leq\lambda\leq 1\}$, which means that half of dephasing channels break entanglement. However, for the two-Pauli channels, $\mathcal{C}_{\mathrm{EBC}}=\{\lambda:\,\lambda=1/2\}$, so this class consists in exactly one entanglement breaking channel with volume zero.

\begin{figure}[htb!]
  \includegraphics[width=0.4\textwidth]{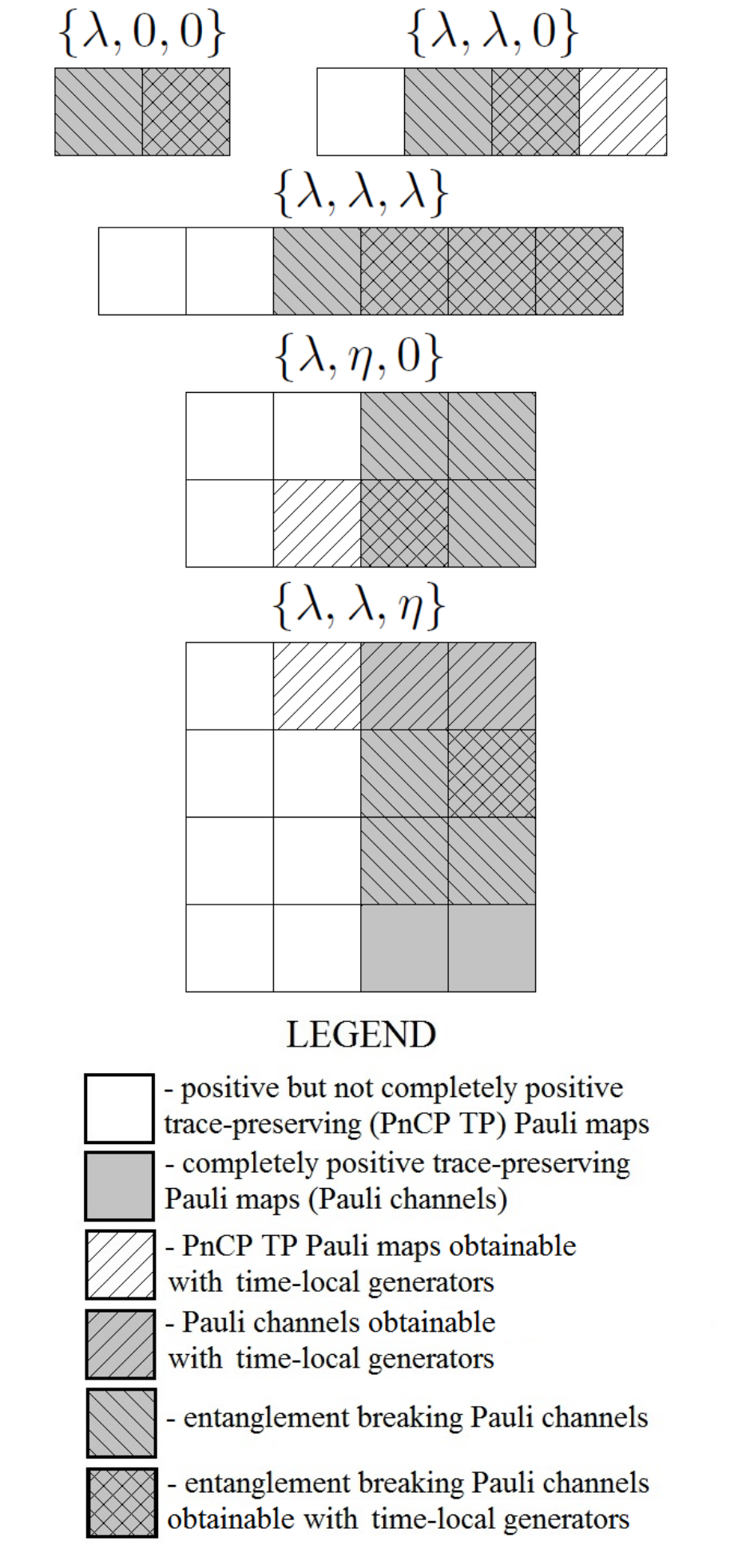}
\caption{
A quantitative representation of the relative volumes for one and two-parameter positive, trace-preserving Pauli maps.}
\label{CPT_chart}
\end{figure}

To summarize, the volume of Pauli maps that are positive and trace-preserving depends on the number of non-zero eigenvalues but not on their degeneracy. Moreover, for two-parameter maps, the relative volumes of Pauli channels and time-local generated positive maps with respect to the positive, trace-preserving maps coincide. For more details, see Fig. \ref{CPT_chart}.

\FloatBarrier

\section{P, CP and L-divisibility}

A quantum channel $\Lambda$ is divisible if it is decomposable into $\Lambda=V\Lambda^\prime$, where $V$ and $\Lambda^\prime$ denote a non-unitary, trace-preserving map and a quantum channel, respectively. If $V$ is positive, then $\Lambda$ is called {\it P-divisible}. Analogically, if $V$ is completely positive, then $\Lambda$ is referred to as {\it CP-divisible}. In addition to P and CP-divisibility, one can introduce an even more restrictive notion of L-divisibility. By definition, $\Lambda$ is L-divisible if and only if there exists $L$ such that $\Lambda=\exp L$ \cite{Wolf}.

In the theory of open quantum systems, the P and CP-divisibility properties of dynamical maps are used to determine non-Markovianity of quantum evolution. Namely, if $\Lambda(t)=V(t,s)\Lambda(s)$ for $s\leq t$ is CP-divisible, then the associated evolution is Markovian according to ref. \cite{RHP}. A PnCP-divisible dynamical map (i.e., P-divisible but not CP-divisible) is called {\it weakly non-Markovian} \cite{witness2,ChManiscalco}, and it describes the dynamics with no information backflow from the system to the environment \cite{Filip2}. Finally, a quantum dynamical map is L-divisible if and only if it a Markovian semigroup \cite{Davalos}.

For the Pauli channels, all the information about divisibility is encoded in their eigenvalues $\lambda_k$ \cite{Cirac}. Namely, the necessary and sufficient condition for P-divisibility reads
\begin{equation}\label{P1}
\lambda_1\lambda_2\lambda_3\geq 0.
\end{equation}
This trivially holds for non-invertible channels, which are also CP-divisible if and only if
\begin{equation}
\exists !_{\,k}\quad\lambda_k\neq 0.
\end{equation}
Meanwhile, invertible Pauli channels are CP-divisible if and only if
\begin{equation}\label{CP1}
0<\lambda_1\lambda_2\lambda_3\leq\lambda_k^2,
\end{equation}
Finally, the property of L-divisibility is equivalent to \cite{Davalos}
\begin{equation}\label{L1}
\lambda_j\lambda_k\leq\lambda_m
\end{equation}
for mutually different $j,k,m$. The associated regions of integration, which lie inside the $\mathcal{C}_{\mathrm{CP}}$ region, are denoted by $\mathcal{C}_{\mathrm{P\mhyphen div}}$, $\mathcal{C}_{\mathrm{CP\mhyphen div}}$, and $\mathcal{C}_{\mathrm{L\mhyphen div}}$, respectively.

Now, let us analyze the divisibility of the Pauli channels considered in the previous section, starting with the one-parameter channels.

\begin{Example}
For the channels with only one non-zero eigenvalue $\lambda$, the P and CP-divisibility conditions coincide. One has
\begin{equation}
V(\mathcal{C}_{\mathrm{P\mhyphen div}})=V(\mathcal{C}_{\mathrm{CP\mhyphen div}})
=V(\mathcal{C}_{\mathrm{CPT}})=1.
\end{equation}
Moreover, every L-divisible channel is time-local generated and vice versa, so they are twice less numerous than CP-divisible channels.
\end{Example}

\begin{Example}\label{Ex1}
Every Pauli channel with two identical non-zero eigenvalues $\lambda$ that is obtainable with a time-local generator is P-divisible, and
\begin{equation}
\frac{V(\mathcal{C}_{\mathrm{P\mhyphen div}})}
{V(\mathcal{C}_{\mathrm{CPT}})}=\frac 12.
\end{equation}
None of these channels are CP or L-divisible, however. For that reason, $V(\mathcal{C}_{\mathrm{CP\mhyphen div}})=V(\mathcal{C}_{\mathrm{L\mhyphen div}})=0$.
\end{Example}

\begin{Example}
For the isotropic channels with all $\lambda_k=\lambda$, the L-divisibility condition simplifies to $0\leq\lambda\leq 1$. Therefore, $\mathcal{C}_{\mathrm{L\mhyphen div}}$ coincides not only with the P and CP-divisibility regions but also with $\mathcal{C}_{\mathrm{CPT}}\cap\mathcal{C}_{\mathrm{TLG}}$. The corresponding volumes are
\begin{equation}
\begin{split}
V(\mathcal{C}_{\mathrm{P\mhyphen div}})
=V(\mathcal{C}_{\mathrm{CP\mhyphen div}})
=V(\mathcal{C}_{\mathrm{L\mhyphen div}})
=&\\V(\mathcal{C}_{\mathrm{CPT}}\cap\mathcal{C}_{\mathrm{TLG}})=&\frac{\sqrt{3}}{2}.
\end{split}
\end{equation}
\end{Example}

Unsurprisingly, one can notice many intersections between the divisibility regions. Next, let us move our attention to the two-parameter channels.

\begin{Example}
The Pauli channels with two different non-zero eigenvalues $\lambda$, $\eta$ always satisfy the P-divisibility condition, and so
\begin{equation}
\frac{V(\mathcal{C}_{\mathrm{P\mhyphen div}})}{V(\mathcal{C}_{\mathrm{CPT}})}=1.
\end{equation}
Similarly to the non-invertible channels from Example \ref{Ex1}, $V(\mathcal{C}_{\mathrm{CP\mhyphen div}})=V(\mathcal{C}_{\mathrm{L\mhyphen div}})=0$.
\end{Example}

\begin{Example}\label{Ex2}
In the last example, we take the channels that have two identical $\lambda$ and one different $\eta$ non-zero eigenvalues. In this case, the integration regions become
\begin{align*}
\mathcal{C}_{\mathrm{P\mhyphen div}}&=\Big\{\lambda,\eta:\,0\leq\eta\leq 1,\,
|\lambda|\leq(\eta+1)/2\Big\},\\
\mathcal{C}_{\mathrm{CP\mhyphen div}}=\mathcal{C}_{\mathrm{L\mhyphen div}}&=
\Big\{\lambda,\eta:\,0\leq\eta\leq 1,\, -\sqrt{\eta}\leq\lambda\leq\sqrt{\eta}\Big\},\\
\mathcal{C}_{\mathrm{P\mhyphen div}}\cap\mathcal{C}_{\mathrm{TLG}}&=
\Big\{\lambda,\eta:\,0\leq\eta\leq 1,\,0\leq\lambda\leq(\eta+1)/2\Big\},\\
\mathcal{C}_{\mathrm{CP\mhyphen div}}\cap\mathcal{C}_{\mathrm{TLG}}&=\\
\mathcal{C}_{\mathrm{L\mhyphen div}}\cap\mathcal{C}_{\mathrm{TLG}}&=
\Big\{\lambda,\eta:\,0\leq\eta\leq 1,\,0\leq\lambda\leq\sqrt{\eta}\Big\}.
\end{align*}
Note that this is the only family of Pauli channels where $\mathcal{C}_{\mathrm{CP\mhyphen div}}=\mathcal{C}_{\mathrm{L\mhyphen div}}\neq\emptyset$.
One finds that
\begin{equation}
\frac{V(\mathcal{C}_{\mathrm{P\mhyphen div}})}{V(\mathcal{C}_{\mathrm{CPT}})}
=\frac 34,
\end{equation}
\begin{equation}
\frac{V(\mathcal{C}_{\mathrm{CP\mhyphen div}})}{V(\mathcal{C}_{\mathrm{CPT}})}
=\frac{V(\mathcal{C}_{\mathrm{L\mhyphen div}})}{V(\mathcal{C}_{\mathrm{CPT}})}
=\frac 23,
\end{equation}
and, for the channels obtainable with time-local generators,
\begin{equation}
\frac{V(\mathcal{C}_{\mathrm{P\mhyphen div}}\cap\mathcal{C}_{\mathrm{TLG}})}{V(\mathcal{C}_{\mathrm{CPT}}\cap\mathcal{C}_{\mathrm{TLG}})}
=1,
\end{equation}
\begin{equation}
\frac{V(\mathcal{C}_{\mathrm{CP\mhyphen div}}\cap\mathcal{C}_{\mathrm{TLG}})}{V(\mathcal{C}_{\mathrm{CP}}\cap\mathcal{C}_{\mathrm{TLG}})}
=\frac{V(\mathcal{C}_{\mathrm{L\mhyphen div}}\cap\mathcal{C}_{\mathrm{TLG}})}{V(\mathcal{C}_{\mathrm{CP}}\cap\mathcal{C}_{\mathrm{TLG}})}
=\frac 89.
\end{equation}
Interestingly, there are relatively more divisible channels of each type in the $\mathcal{C}_{\mathrm{TLG}}$ region than outside of it.
\end{Example}

\FloatBarrier

\begin{figure}[htb!]
  \includegraphics[width=0.4\textwidth]{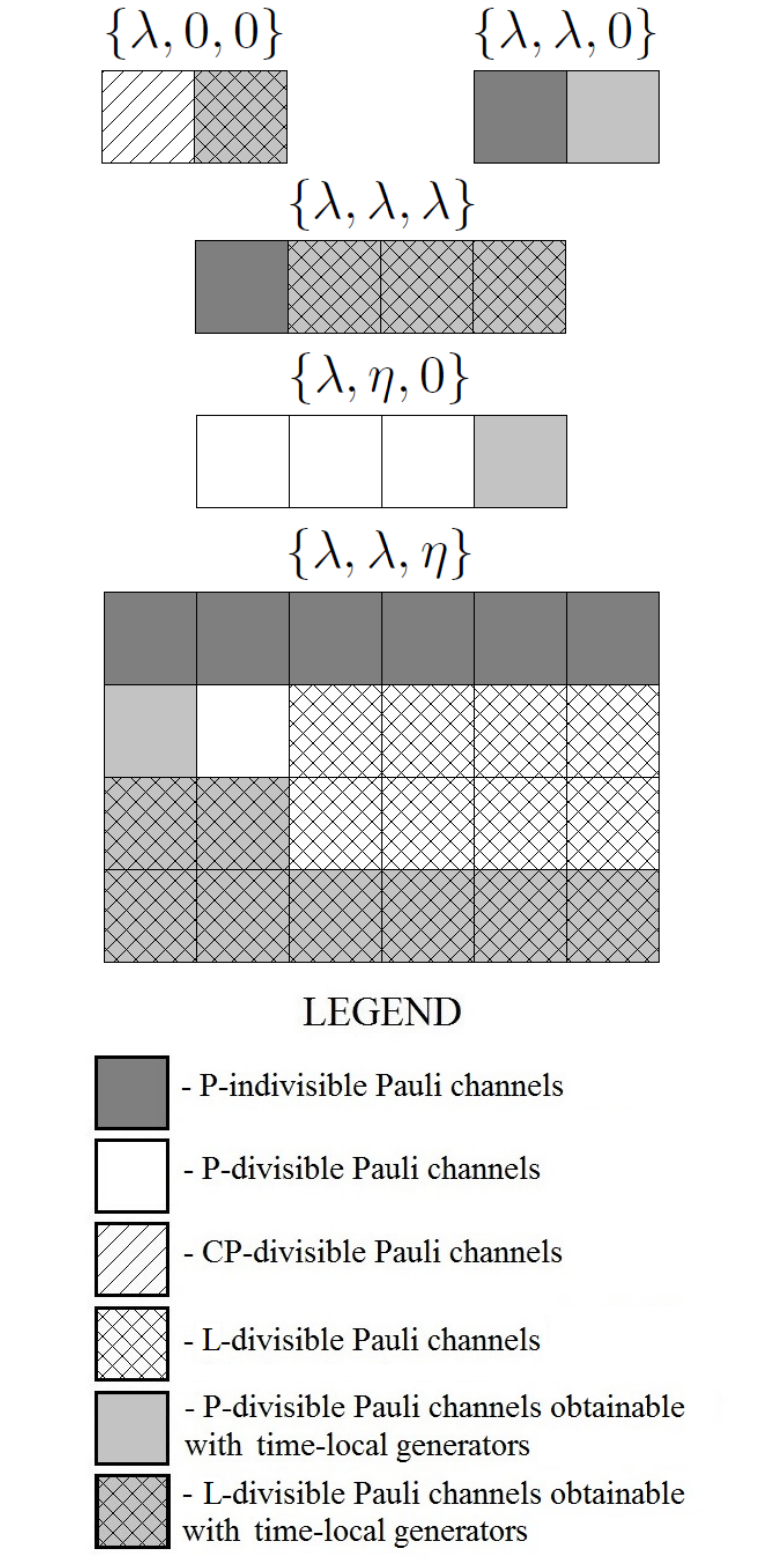}
\caption{
A quantitative representation of the relative volumes for one and two-parameter Pauli channels.}
\label{div_chart}
\end{figure}

For the completeness sake, we observe that every two-Pauli channel is PnCP-divisible, whereas all dephasing channels are L-divisible. Moreover, for the Pauli channels with three distinct eigenvalues, one has $\mathcal{C}_{\mathrm{L \mhyphen div}}=\mathcal{C}_{\mathrm{CP \mhyphen div}}\cap\mathcal{C}_{\mathrm{TLG}}$.

In conclusion, we observe that the CP and L-divisibility regions coincide for all channels considered in Examples 2--5. If the Pauli channel has two non-zero eigenvalues, then $V(\mathcal{C}_{\mathrm{CP\mhyphen div}})=V(\mathcal{C}_{\mathrm{L\mhyphen div}})=0$. This result agrees with ref. \cite{Sagnik}, where the authors showed that if a qubit map has exactly one vanishing eigenvalue, then it is not CP-divisible. Additionally, the time-local generated channels that are not P-divisible form a null set, as
\begin{equation}
\frac{V(\mathcal{C}_{\mathrm{P\mhyphen div}}\cap\mathcal{C}_{\mathrm{TLG}})}
{V(\mathcal{C}_{\mathrm{TLG}})}=1.
\end{equation}
Every one-parameter channel with three identical eigenvalues that is P-divisible is automatically L-divisible, as well. For a graphical illustration of the relative volumes, refer to Fig. \ref{div_chart}.

\FloatBarrier

\section{Conclusions}

In this paper, we analyze the geometry of positive, trace-preserving Pauli maps and Pauli channels with degenerated or vanishing eigenvalues. In particular, we analytically calculate their volumes using the Hilbert-Schmidt volume element of the corresponding Choi-Jamio{\l}kowski states. Next, we find the relative volumes for special subclasses of the Pauli channels, including the maps that are divisible or obtainable with time-local generators.

Our results can be interpreted as the probability that the Pauli channel that belongs to a specific class has certain properties. Interestingly, even if we know nothing about the channel eigenvalues, we still possess some partial knowledge about the relative volumes. The time-local generated channels that are not P-divisible have volume zero, so they form at most a null set. Also, the volumes of CP and L-divisible channels with positive eigenvalues for channels with two non-zero eigenvalues are zero.

In further work, one can consider the geometry of other quantum channels, like amplitude damping channels or generalizations of Pauli channels. However, working with higher-dimensional systems may require some numerical calculations.  An interesting open problem is to establish whether the channel volumes depend on the choice of the metric. Comparing the shape and size of the Choi-Jamio{\l}kowski states in the space of all quantum states can shed some light on the structure of quantum channels. Another task would be to analyze how the relative volumes of dynamical maps change in time under legitimate quantum evolution.

\section*{Acknowledgements}

This paper was supported by the Polish National Science Centre project No. 2018/31/N/ST2/00250.

\bibliography{C:/Users/cynda/OneDrive/Fizyka/bibliography}
\bibliographystyle{C:/Users/cynda/OneDrive/Fizyka/beztytulow2}

\end{document}